\documentclass[5p,authoryear]{elsarticle}
\usepackage[utf8]{inputenc}
\usepackage{booktabs}
\usepackage{tabularx}
\usepackage{ragged2e}
\usepackage[table]{xcolor}
\usepackage{natbib}
\usepackage{fancyhdr}
\usepackage{hyperref}
\hypersetup{
    colorlinks=true,   
    urlcolor=blue,
    citecolor=blue,
    linkcolor=blue
}

\newcommand{\apj}{ApJ}
\newcommand{\apjl}{ApJ Lett.}
\newcommand{\apjs}{ApJ Suppl.}
\newcommand{\aj}{AJ}
\newcommand{\mnras}{MNRAS}
\newcommand{\aap}{A\&A}

\makeatletter
\def\blfootnote{\gdef\@thefnmark{}\@footnotetext}
\makeatother

\newcommand{\dmo}{dark-matter-only}
\newcommand{\vmax}{v_{\rm max}}
\newcommand{\Msun}{\,\mathrm{M}_{\odot}}

\begin{document}
\begin{frontmatter}

\title{Improving Performance of Zoom-In Cosmological Simulations using Initial Conditions with Customized Grids}

\author{Gillen Brown\href{https://orcid.org/0000-0002-9114-5197}{\includegraphics[scale=0.5]{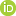}}}\corref{cor1}
\ead{gillenb@umich.edu}

\author{Oleg Y. Gnedin\href{https://orcid.org/0000-0001-9852-9954}{\includegraphics[scale=0.5]{orcid.png}}}

\address{Department of Astronomy, University of Michigan, Ann Arbor, MI 48109, USA}
\cortext[cor1]{Corresponding Author}

\begin{abstract}
We present a method for customizing the root grid of zoom-in initial conditions used for simulations of galaxy formation. Starting from the white noise used to seed the structures of an existing initial condition, we cut out a smaller region of interest and use this trimmed white noise cube to create a new root grid. This new root grid contains similar structures as the original, but allows for a smaller box volume and different grid resolution that can be tuned to best suit a given simulation code. To minimally disturb the zoom region, the dark matter particles and gas cells from the original zoom region are placed within the new root grid, with no modification other than a bulk velocity offset to match the systemic velocity of the corresponding region in the new root grid. We validate this method using a zoom-in initial condition containing a Local Group analog. We run collisionless simulations using the original and modified initial conditions, finding good agreement. The dark matter halo masses of the two most massive galaxies at $z=0$ match the original to within 15\%. The times and masses of major mergers are reproduced well, as are the full dark matter accretion histories. While we do not reproduce specific satellite galaxies found in the original simulation, we obtain qualitative agreement in the distributions of the maximum circular velocity and the distance from the central galaxy. We also examine the runtime speedup provided by this method for full hydrodynamic simulations with the ART code. We find that reducing the root grid cell size improves performance, but the increased particle and cell numbers can negate some of the gain. We test several realizations, with our best runs achieving a speedup of nearly a factor of two. 
\end{abstract}

\begin{keyword}
galaxies: formation \sep methods: numerical
\end{keyword}
\end{frontmatter}

\section{Motivation}

Numerical simulations are an indispensable tool for understanding the complex processes that occur during galaxy formation. Various groups, using different numerical approaches, have used simulations to successfully replicate galaxy properties and interpret observations \citep[e.g.,][]{vogelsberger_etal20}. One approach to simulating individual galaxies within a cosmological context is with so-called "zoom-in" initial conditions \citep{navarro_etal94, hahn_etal11_music}. They rely on first simulating a large volume of the universe containing many galaxies with coarse resolution, and selecting a region around a particular galaxy, such as one resembling the Milky Way. Then this comparatively small region is resampled with many more resolution elements while keeping the initial low resolution in the rest of the volume. The zoom-in technique allows simulations to model galaxies specifically picked to have desirable properties at very high resolution, enabling detailed modeling of the relevant physical processes. \blfootnote{\copyright \ 2020. This manuscript version is made available under the \href{http://creativecommons.org/licenses/by-nc-nd/4.0/}{ CC-BY-NC-ND 4.0 license}.}

Particularly common are zoom-in simulations designed to target the Milky Way or larger Local Group volume. Recent surveys \citep[e.g.][]{gaia_dr2, Majewski_eta17_apogee} have collected a large amount of data for Milky Way and other Local Group galaxies. To help interpret these observations, theorists have long been simulating isolated galaxies with similar masses to the Milky Way \citep[e.g.,][]{hopkins_etal14}. However, recent advances both in the knowledge of the Local Group's formation history \citep{hammer_etal07, dsouza_bell18} and improved understanding from simulations of how isolated galaxies are different from those in groups \citep{santistevan_etal20}, have shown that the Local Group's environment is essential to understanding its formation. As a result, any computational setup aiming to uncover the origin of the Milky Way must be specifically tailored to match the known properties of the full Local Group, including the assembly histories of its galaxies. 

As several key galaxy properties such as halo mass and merger history are determined by the initial conditions (ICs) used in a given simulation, these ICs play a critical role in any comparison of results between different simulations. Many galaxy formation simulations have been run by different groups, using different codes and different ICs  \citep[e.g.,][]{hopkins_etal14,ceverino_etal14,vogelsberger_etal14_illustris,wang_etal15_nihao,schaye_etal_15_eagle,sawala_etal16_apostle,wetzel_etal16_fire_latte,grand_etal_17_auriga,li_etal17,hopkins_etal18_fire2}. These different ICs can lead to different galaxy properties, even before accounting for differences in modeled physics or numerical implementation, making a comparison of results challenging. If groups were to use the same ICs (particularly for zoom-in ICs that replicate the properties of the entire Local Group), any differences caused by different ICs would be eliminated, making it easier to understand how modeling differences affect the results. However, such zoom-in simulations are often computationally expensive, requiring on the order of $10^6$ CPU hours  \citep{wetzel_etal16_fire_latte, hopkins_etal18_fire2}. Importantly, the performance of different codes may be affected by the details of the ICs, particularly for zoom-in ICs, where the root grid cells are usually coarse and most of the volume of the box is uninteresting. This may make simulations using some ICs prohibitively expensive when run using certain codes. Customizing these parameters, while preserving the properties of the zoom region, may improve computational efficiency and lower the cost of these high-resolution simulations of galaxy formation, making a commonly used set of initial conditions more practical and facilitating code comparison.

Here we present a method of customizing the root grid in ICs to improve code performance. This method can be used to reduce the box size and increase the resolution of the root grid. As an example we use the ``Thelma \& Louise" IC, initially presented in \citet{garrison-kimmel_etal14_elvis}. It is a zoom-in IC that contains a Local Group analog. In particular, the merger history of the Milky Way and M31 analogs has a qualitative resemblance to the Local Group. 

We validate the method using the Adaptive Refinement Tree (ART) code \citep{kravtsov_etal97,kravtsov99,kravtsov03,rudd_etal08}. The ART code uses adaptive mesh refinement of the simulation grid to provide high resolution only in interesting regions where it is needed. Specifically, it starts with a uniform grid of root cells that are refined if they meet one of several criteria, such as a density threshold or a comparison to the local Jeans length. The size of the root grid cells is relevant because the load balancing algorithm operates on these root cells. Each root grid cell and all its refined ``children" are assigned to a given MPI rank. Very large root grid cells therefore may contain entire structures that are all assigned to a given rank. For example, in a Local Group-like simulation, both main galaxies may each be located in single root grid cells, with very little in any other root grid cells. This decreases efficiency for two reasons. First, it is difficult to evenly divide the needed workload among different computational nodes. Such uneven load balancing may result in one rank taking much longer than the rest, wasting computation time as other processors wait. Second, it is difficult for the ART code to scale to a large number of nodes, as adding more nodes will not bring any benefit if the work cannot be subdivided that finely. Reducing the root grid cell size will address both of these issues.

We also demonstrate that this method reproduces basic properties of the target galaxies, such as their $z=0$ dark matter halo masses, growth histories (including times and masses of major mergers), and their populations of satellite galaxies.

\begin{figure}
    \includegraphics[width=0.483\textwidth]{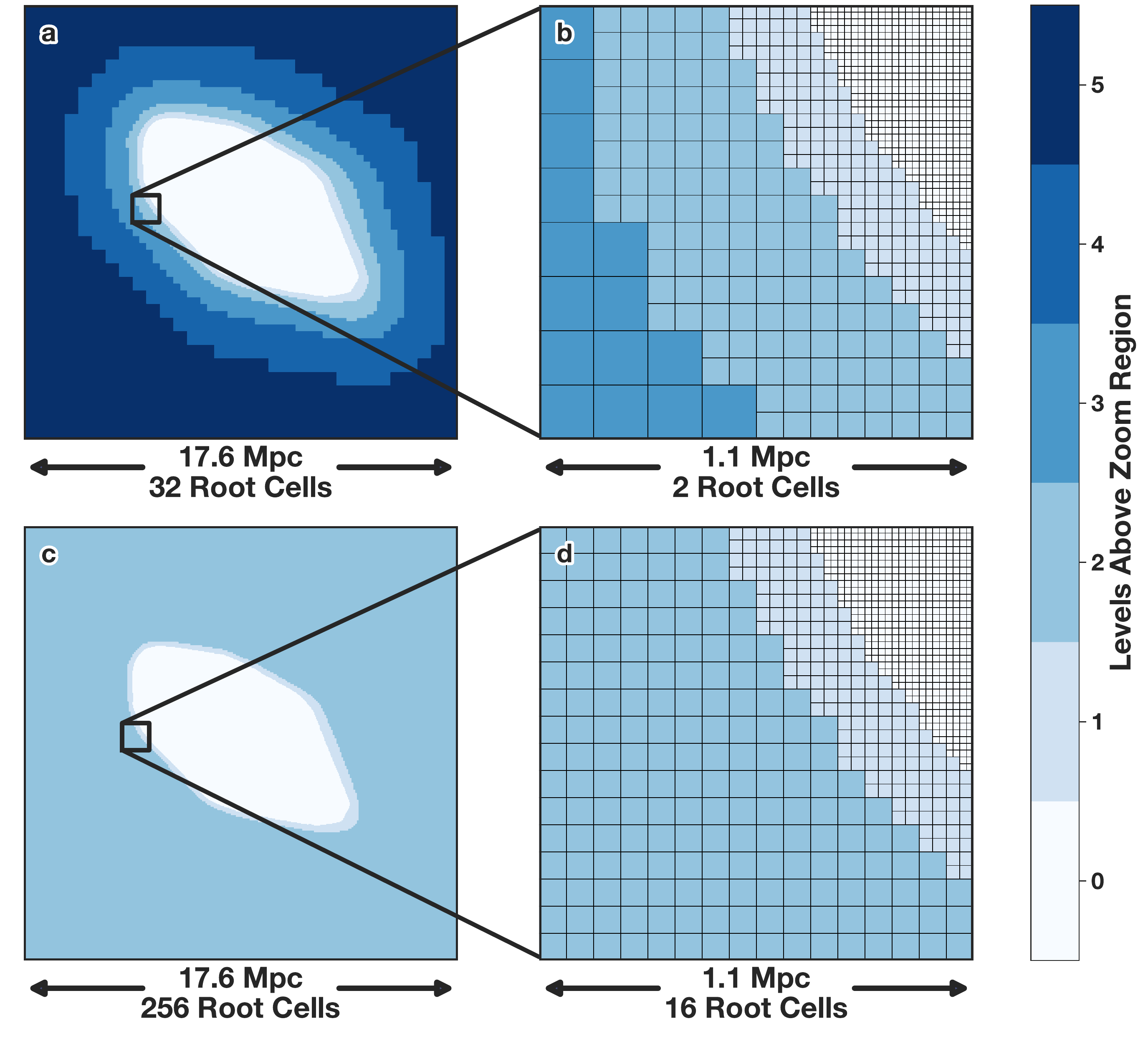}
    \vspace{-5mm}
    \caption{An illustration of how the grid structure changes when using our method. Panel \textbf{(a)} shows the level of refinement in the original IC. The maximum zoom level contains all the particles that end up in the main galaxies at $z=0$. Note that this is an inset of the original box, which is 70.4 Mpc across. Panel \textbf{(b)} shows a further inset, with cell boundaries outlined. Panels \textbf{(c)} and \textbf{(d)} are the equivalents from the modified IC with a box length of 17.6 Mpc and a $256^3$ root grid. Note how the original root grid and lower-resolution intermediate levels are now replaced by a new higher-resolution root grid, while the original zoom region and high resolution intermediate levels are preserved.}    
    \label{fig:grid_demo}
\end{figure}

\begin{figure*}  
    \centering 
    \includegraphics[width=0.77\textwidth]{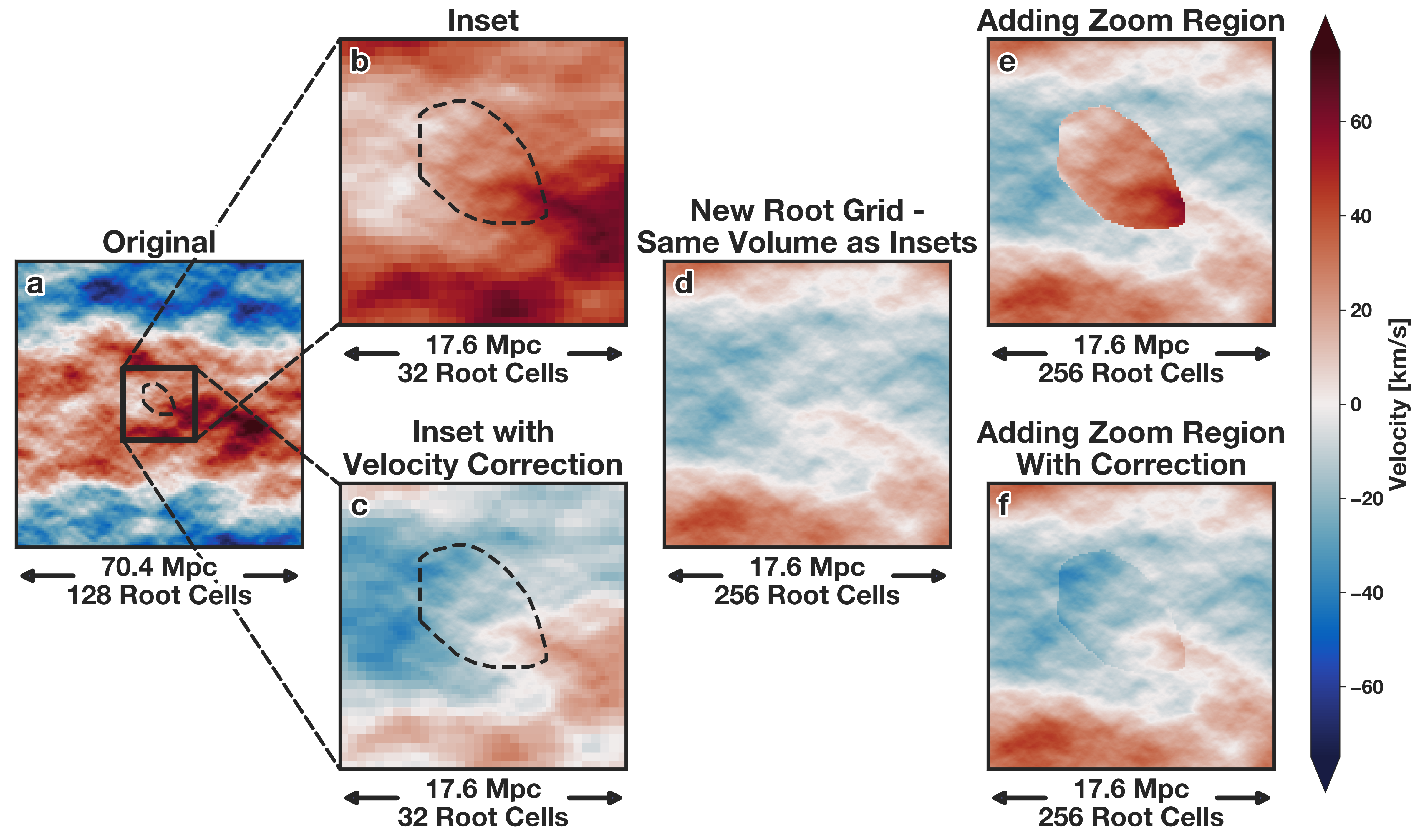}
    \vspace{-2mm}
    \caption{An illustration of the process of customizing ICs. All panels show the $z$-axis component of the gas velocity in the same slice through the simulation volume, with a dashed black line outlining the zoom region. Panel \textbf{(a)} shows the original realization, while panel \textbf{(b)} shows an inset of the region that will become the new full simulation volume. Note the higher resolution cells in the central refined region, with low resolution cells in the outskirts. This region has a nonzero bulk velocity, so we subtract it off to obtain panel \textbf{(c)}. Panel \textbf{(d)} shows the new root grid constructed with our method, which represents the same volume of space as the insets \textbf{(b, c)}. A comparison with panel \textbf{(c)} shows that many of the key features are present in both panels, with differences near the edges. These are required for periodic boundary conditions. Note that in panel \textbf{(d)} the velocity at the top boundary matches the bottom, and left matches right, unlike panel \textbf{(c)}. The higher resolution root grid cells are also apparent in the new root grid, as the outskirts do not have the pixelation present in panels \textbf{(b)} and \textbf{(c)}. Panels \textbf{(e)} and \textbf{(f)} show the result of combining the new root grid with the zoom region from panels \textbf{(b)} and \textbf{(c)} respectively. The velocity correction is necessary to give a smooth velocity field. Panel \textbf{(f)} is used as our final IC.} \label{fig:cut_demo}
\end{figure*}

\section{Method}

A detailed description of the process of creating our cosmological ICs can be found in \citet{hahn_etal11_music}, but we summarize some of the key points here to provide context for our method. To start, a cube of a chosen cosmic volume is covered by a uniform 3D grid, known as the ``root grid." A white noise field is obtained by assigning a random value sampled from a $\mathcal{N}(0,\,1)$ distribution to each cell in the root grid. This is then convolved with the matter power spectrum to obtain realistic matter overdensities. The convolution is done using a Fast Fourier Transform, which enforces periodic boundary conditions on the cube. Lagrangian perturbation theory \citep[e.g.][]{zeldovich_70} is used to turn the overdensity field into particle positions and velocities. When used for grid-based codes such as ART, this process creates the gas densities and velocities for root grid cells, as well as the positions and velocities for each cell's corresponding dark matter particle (``root grid particle").

When producing zoom-in ICs, a region of interest is selected, then a finer grid is used in that region. The spacing of the finer grid can be arbitrarily small, only limited by the computational resources required to run a simulation at such high resolution. A buffer of intermediate grid levels is used around the zoom region to transition smoothly to unrefined regions. The process for generating the properties of the dark matter particles and gas cells (referred to as ``zoom particles" and ``zoom cells") in the zoom region follows the same basic process as the root grid, just on a finer grid and with constraints to ensure consistency with the root grid (see \citealt{hahn_etal11_music} for full details). 

Our goal is to customize the root grid while preserving the environment of the zoom-in region. To achieve this, we create a hybrid IC, where we produce a new root grid that uses a smaller box size with a higher resolution grid, then embed the zoom region from the original IC. Keeping the original zoom region intact ensures that the galaxies in this region are minimally disturbed.

Creating the new root grid starts by taking the white noise field of the original cube and cutting out a smaller white noise cube. The same process (convolution with the power spectrum and gravity calculations) is used to create a new IC from this trimmed white noise cube. As the white noise cube is what seeds the resulting structures, they will be nearly preserved in the new box. The only changes will be due to the enforcement of periodic boundary conditions by the Fast Fourier Transform. This process results in root grid cells and particles with properties nearly identical to those in the original, with only slight modifications near the boundaries (far from the region of interest) to ensure periodicity. The resulting root grid can be regenerated at any desired resolution. This allows us to have significantly smaller root grid cells, potentially improving performance in codes like ART that use root grid cells for load balancing.

After creating the new root grid, we combine it with the particles and gas cells from the original zoom region. We start by keeping all the particles and cells from the zoom region of the original IC, as we want to minimally disturb the zoom region. We also keep the particles and cells from the intermediate levels that are at the same level or deeper than the new root grid. We then go through all the new root grid particles and remove any that are in the same root grid cell as a previously kept particle. Figure~\ref{fig:grid_demo} shows an example of the new grid structure.

\begin{table*}
    \centering
    \renewcommand{\arraystretch}{1.2}  
    \rowcolors{2}{white}{gray!15} 
    \begin{tabular}{llccc}
        \toprule
        Run & Box Length [Mpc] & Number of Root Grid Cells & Root Grid Cell Size [kpc]  & Number of Particles \\
        \midrule
        L1-128 & 70.4 (Original) & $128^3$ & 550.0\phantom{0} & 51,196,181 \\
        L2-256 & 35.2 & $256^3$ & 137.5\phantom{0} & 65,589,360 \\
        L4-128 & 17.6 & $128^3$ & 137.5\phantom{0} & 50,908,382 \\
        L4-256 & 17.6 & $256^3$ & \phantom{0}68.75 & 64,538,268 \\
        \bottomrule
    \end{tabular}
    \caption{List of initial conditions used to run \dmo simulations. Note that all realizations have the same zoom region that includes 47,517,792 particles, which dominates the total count.}
    \label{tab:sims}
\end{table*}

Importantly, we also need to correct the velocities of the zoom particles. The new root grid will have a mean velocity of zero by construction, as the simulation box is not moving in any particular direction, while the equivalent section of the original volume may have some bulk velocity. We calculate the mean velocity of the root grid particles that are replaced, and correct the mean velocity of the zoom particles to match. This important correction minimizes the discontinuity between the zoom region and the root grid and is illustrated in Figure~\ref{fig:cut_demo}.

This simple method, while designed to minimize the discontinuity between the zoom region and root grid, nevertheless cannot fully eliminate the discontinuity. However, this approach allows us to preserve the properties of the zoom region as much as possible (other than its bulk velocity). It results in $z=0$ galaxy properties most similar to the original. In addition, since there are intermediate levels between the maximally refined zoom region and the root grid, the discontinuities will be far from the galaxies of interest. As the zoom region is typically selected to include all particles that end up within the virial radius of the main galaxies, the particles far outside do not play a strong role in the evolution of the galaxies. This minimizes the affect of velocity discontinuity.

We illustrate our method by generating several modified versions of the  Thelma \& Louise IC. The original IC is a 70.4 comoving Mpc box with a $128^3$ root grid \citep{garrison-kimmel_etal14_elvis}. A non-spherical zoom region is approximately 10 comoving Mpc across. This zoom region is 5 levels below the root grid, with intermediate levels nested around the zoom region (see Figure~\ref{fig:grid_demo} for the grid structure in this IC). We create modified versions where we decrease the box size by factors of 2 and 4. A smaller box should improve the simulation runtime, but may cause too much disruption to the regions near the galaxies of interest, affecting their evolution \citep{neyrinck_etal04}. We also use two options for the number of root grid cells, as it is a primary driver of ART's load balancing. Table~\ref{tab:sims} details the suite of ICs we generated, including abbreviated names for each realization, which we use throughout the rest of this paper.

\section{Comparison of Galaxy Properties}

To verify the success of this method, we now evaluate how well the properties of the galaxies at the present time are preserved. To do this, we use all the ICs detailed in Table~\ref{tab:sims} to run \dmo simulations. We use the \texttt{ROCKSTAR} halo finder \citep{behroozi_etal13_rockstar} and the \texttt{Consistent Trees} code \citep{Behroozi_etal13_trees} to generate halo catalogs and merger trees. 

\begin{figure}
    \centering 
    \includegraphics[width=0.483\textwidth]{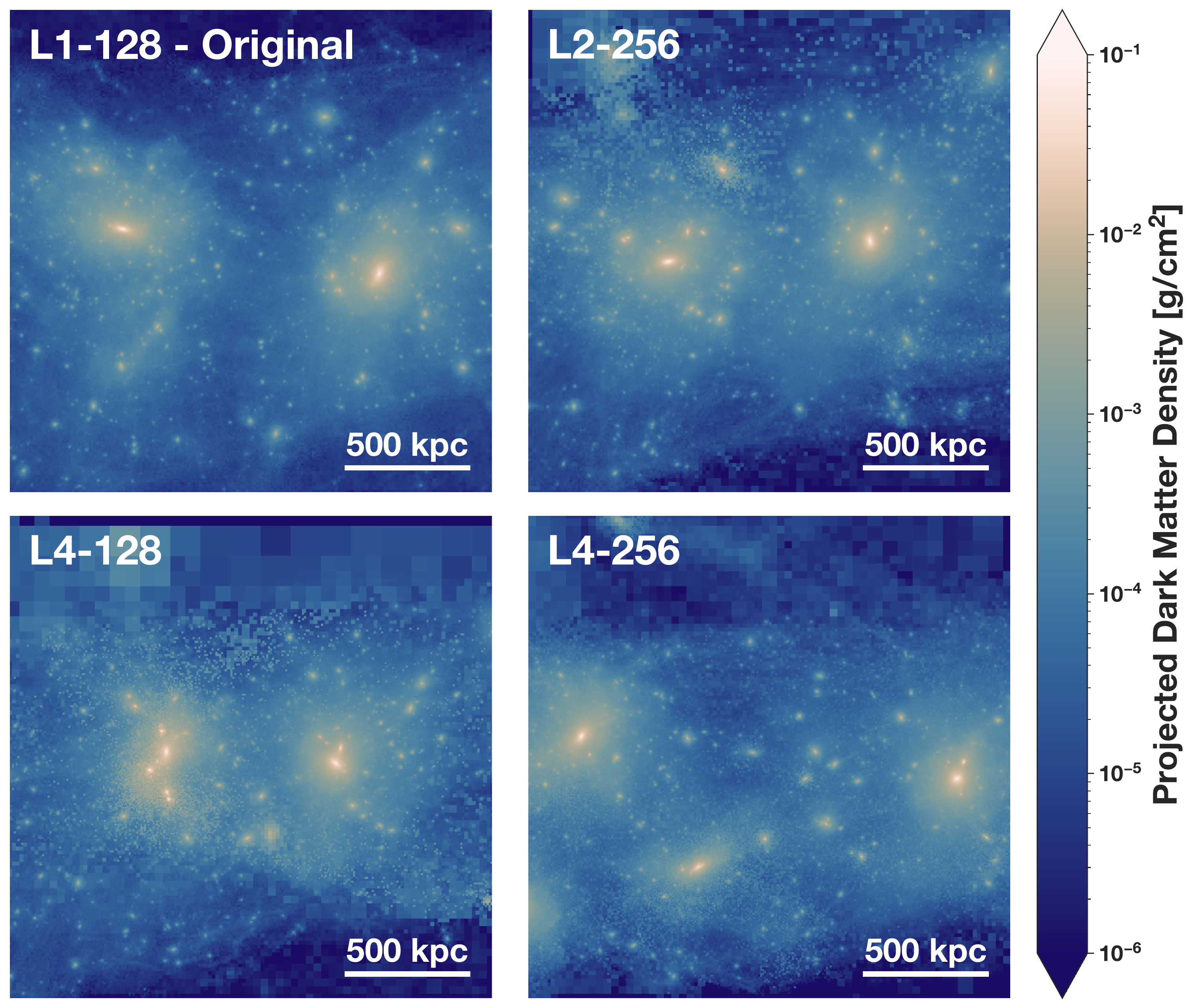}
    \vspace{-5mm}
    \caption{Projected dark matter density of the region surrounding the Local Group analogs at $z=0$ in our dark matter only simulation runs.} 
    \label{fig:projection}
\end{figure}

As a visual representation of the simulations, Figure~\ref{fig:projection} shows the projected dark matter density of the region surrounding the two main galaxies. Their halos are clearly visible in similar positions, with similar large scale filamentary structures.

\begin{figure*}  
    \centering 
    \includegraphics[width=0.8\textwidth]{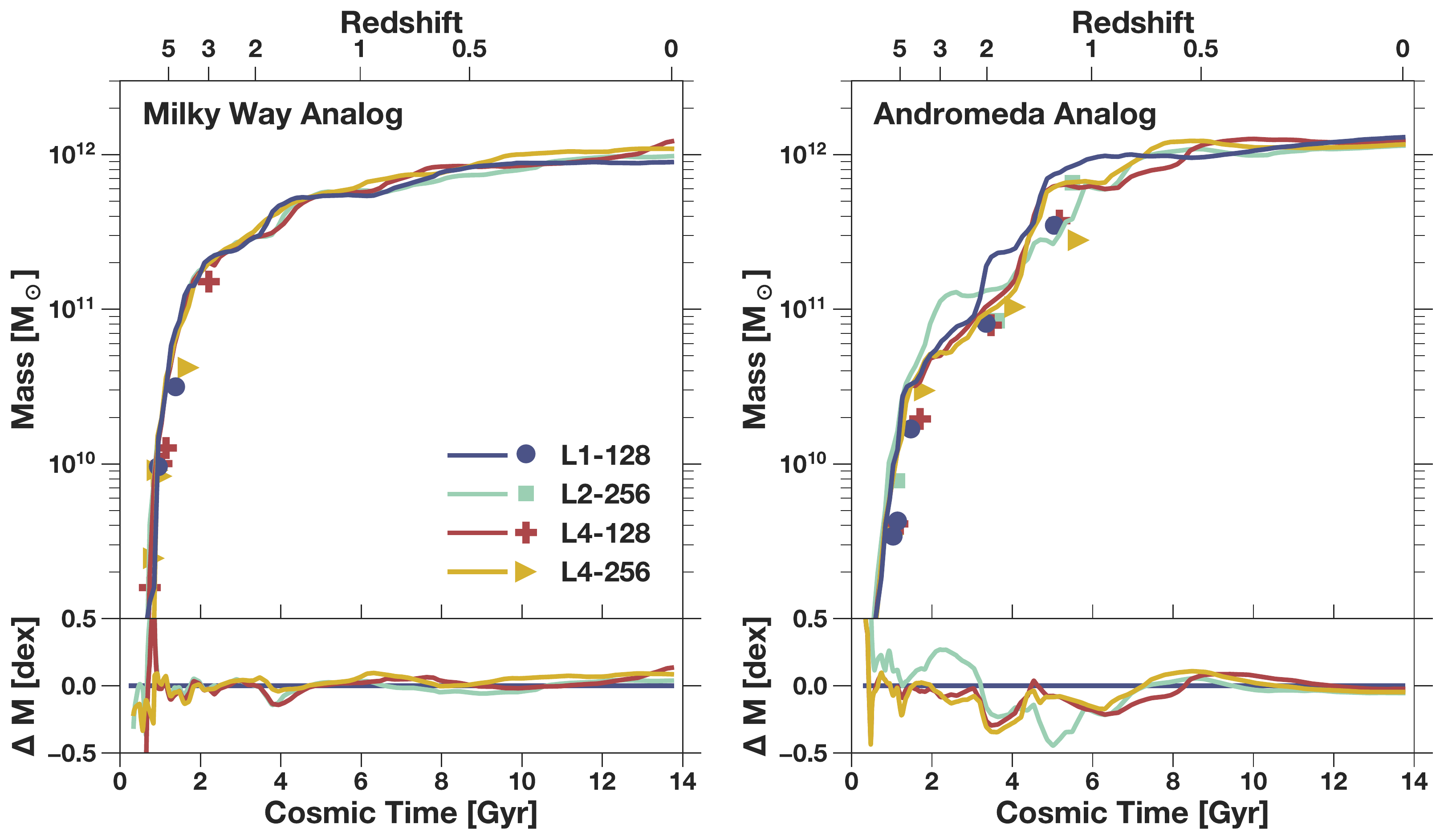}
    \vspace{-2mm}
    \caption{The growth of the dark matter halos of the two most massive galaxies in the Local Group. Symbols show when significant galaxy mergers occur, and are placed at the time of the merger and the mass of the infalling satellite. The bottom panels show the differences in mass as a function of time. Note the excellent agreement at all times in the Milky Way analog, and at late times in the Andromeda analog. Larger differences can occur near the times of mergers, but the final agreement is excellent.}
    \label{fig:halo_growth}
\end{figure*}

As the original IC was chosen because of its close match to the observed present-day mass of the Milky Way and Andromeda galaxies, our simulations using modified ICs must reproduce these final masses and growth histories. Figure~\ref{fig:halo_growth} shows the mass assembly history for the analogs of the Milky Way and Andromeda for different simulation runs. The Milky Way analog has a smooth growth history with few mergers \citep{hammer_etal07}, while the Andromeda analog has a more violent accretion history \citep{dsouza_bell18}, qualitatively matching observations and making this IC a good representation of the Local Group. The final halo masses are reproduced well in all simulations. The $z=0$ halo masses for the Andromeda analog are all within 15\% or the original, while the largest difference in the Milky Way analog is 30\%, due to late growth in the L4-128 run. These halos are all still comparable to the real Milky Way and M31, making them useful for studies of these galaxies.

While galactic mergers are extremely common in the first few billion years of the universe's history, later mergers can make a significant impact on galactic stellar content, and so are more important to reproduce. The simulated Milky Way has no significant mergers after about 2 Gyr in any realization. The Andromeda analog does have a few major mergers, which are reproduced well. Note the clumps of points at around 4 and 6 Gyr in Figure~\ref{fig:halo_growth}. These are two mergers that occur at roughly the same time in all simulations.

The merger histories of the two galaxies are also reflected in their growth histories. The Milky Way analog's quieter accretion history results in smooth growth that is reproduced extremely well, with typical deviations of less than 10\%. Some larger deviations are driven by differences in timing, for example the more rapid growth at $z\approx2$ that occurs later in the L4-128 run than in the original. The Andromeda analog shows larger variations, driven primarily by its more violent accretion history. Large differences in mass (up to nearly a factor of 2) can occur around the times of these major mergers, primarily driven in the differences in timing of rapid growth events. Once the mergers are completed, though, the agreement is excellent, with typical deviations of 10\% after $z=0.5$.

\begin{figure*}  
    \centering 
    \includegraphics[width=1.0\textwidth]{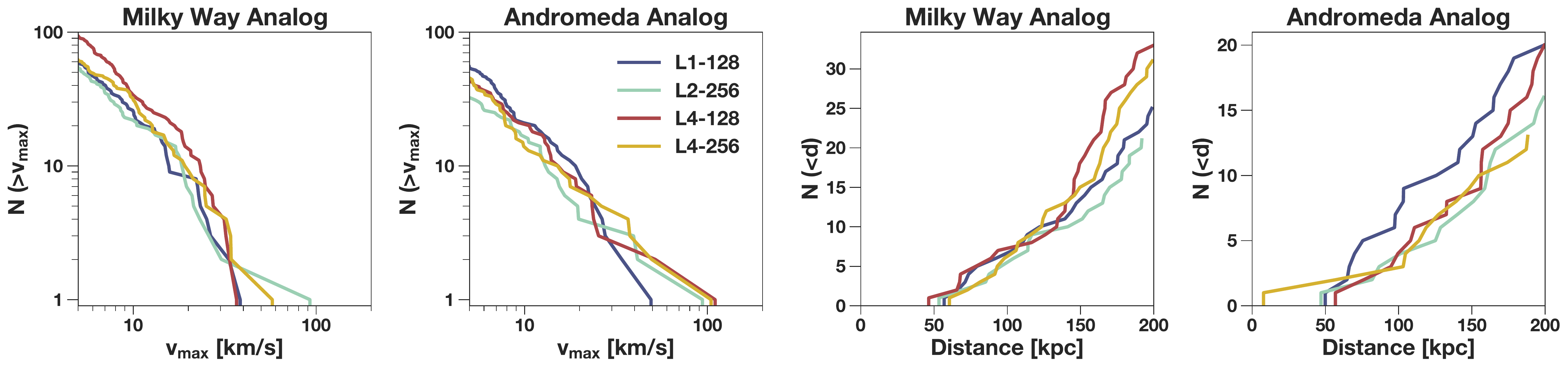}
    \vspace{-6mm}
    \caption{Properties of the satellite galaxies within 200 kpc of the two central galaxies at $z=0$. The left two panels show the cumulative distribution of the maximum circular velocity, while the right two panels show the cumulative radial distribution of satellites with maximum circular velocity above 10~km~s$^{-1}$ (most likely to produce observable stars).}
    \label{fig:satellites}
\end{figure*}

In addition to the central Milky Way and Andromeda galaxies, each has many smaller satellite galaxies. We examine the properties of all dark matter subhalos within 200 kpc (approximately the virial radius) of the two main halos. The left panels of Figure~\ref{fig:satellites} show the distribution in the maximum circular velocities ($\vmax$) of the satellites. This quantity is calculated by \texttt{ROCKSTAR} using the dark matter particles identified as belonging to a given satellite. This makes $\vmax$ subject to numerical discreteness, making exact replication difficult. While the distributions follow similar shapes, the normalization can differ between different runs. One other noteworthy difference is the mass of the most massive satellite, where most modified realizations have a largest satellite significantly more massive than the original run. Interestingly, this larger circular velocity agrees better with the measured circular velocities of the Large Magellanic Cloud around the Milky Way \citep{van_der_marel_etal14_lmc} and M33 around Andromeda \citep{corbelli_03_m33}.

The positions of satellites relative to the central galaxies are also important. As the trajectory of a particle is a solution to systems of partial differential equations, small differences separate exponentially over time. This leads to chaotic non-linear orbital dynamics, making it difficult to obtain exactly the same positions relative to the galaxy center. In the Aquarius suite of simulations, positions of satellites were reproduced very well when changing the resolution of the ICs, showing that positions can be preserved after small changes to the ICs \citep{springel_08_aquarius}. Unlike Aquarius, our realizations change the large scale structures around the Local Group analog, resulting in changes to the positions of the main galaxies, as well as significant changes to the positions of the satellites around each central (see the visual differences apparent in Figure~\ref{fig:projection}). However, we find that the radial distributions of satellites are similar. The right panels of Figure~\ref{fig:satellites} show the cumulative radial distribution of satellite galaxies of the Milky Way and Andromeda analogs with $\vmax > 10$~km/s (which selects halos likely to produce observable stars). For the Milky Way analog, the agreement is excellent. The radial distributions of satellites agree within 150 kpc. Beyond that, the modified realizations start to diverge, reflecting differences in the total number of satellites with $\vmax > 10$~km/s. In the Andromeda analog, while the original realization does have significantly more satellites closer in, the shapes of the distributions are quite similar, with differences in normalization again due to differences in the total number of satellites. Note that in this comparison we only considered satellites within 200~kpc, and for the radial distribution plots chose a cut at 10~km/s in circular velocity. Changing these cuts would change the details of all distributions, but does not change the qualitative agreement present in all panels.

\begin{figure*}  
    \centering 
    \includegraphics[width=\textwidth]{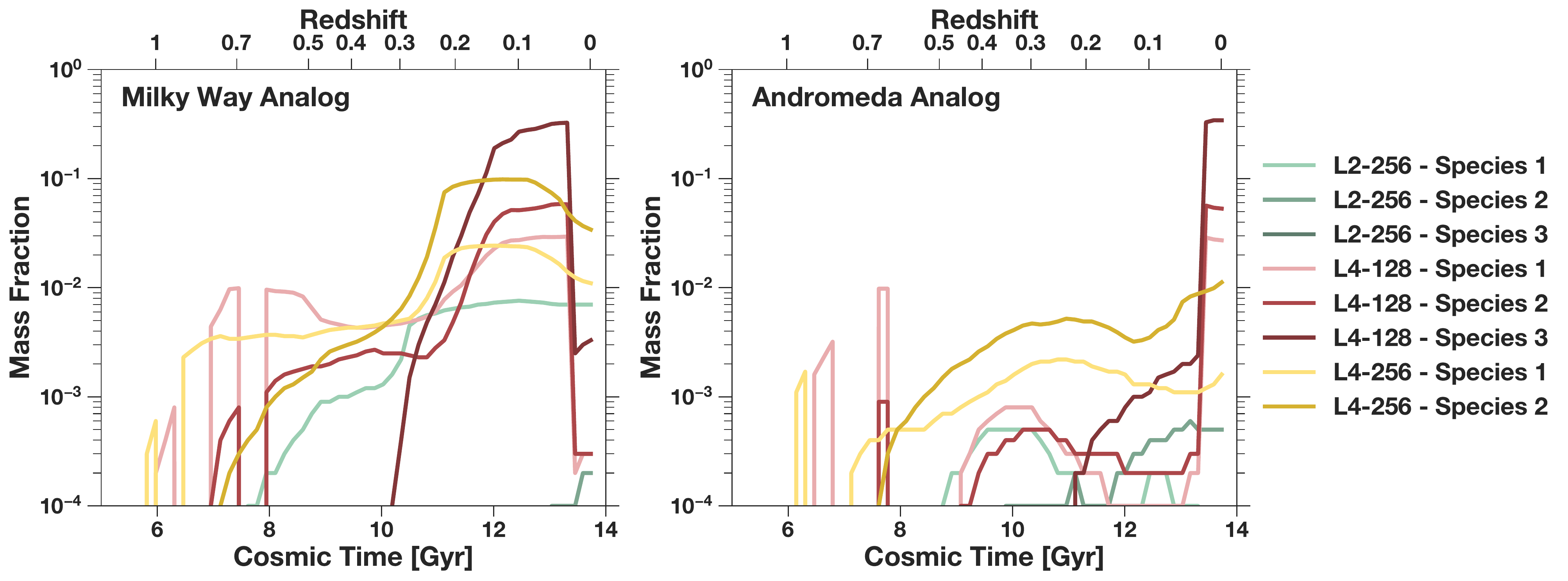}
    \vspace{-5mm}
    \caption{The fraction of mass within the virial radii of the two most massive galaxies contributed by particles other than the zoom particles. Not shown is Species 0, the most highly refined zoom particles, which constitute most of the mass. Higher species number corresponds to higher particle mass. The dramatic contamination in L4-128 run comes from several satellites that fly by the Milky Way analog to the Andromeda analog.}
    \label{fig:contamination}
\end{figure*}

As our method subtly modifies the large scale structure around the zoom region, it may change which particles end up in the final $z=0$ halos. While contamination from higher mass particles was not identically zero in the runs using modified ICs, in most runs it remained insignificant. Figure~\ref{fig:contamination} shows the fraction of mass within the virial radii of the two central galaxies contributed by particles other than the zoom particles. We found higher contamination in our smallest boxes, indicating that moving the periodic boundary conditions closer to the target galaxy modified the structure more strongly. In the L4-128 run, a satellite made entirely of root grid particles flew by both two central galaxies at $z\approx0.1$, while in the L4-256 run a smaller root grid satellite came near the Milky Way analog. The L2-256 run had less contamination, with no infalling satellites made of root grid particles. However, about 1\% of the virial mass of the Milky Way analog came from intermediate level particles. By identifying the contaminating particles in all runs and tracking them back to the IC, we determined that all contaminating particles came from just outside the zoom region. Regenerating the zoom region within the original IC to include the region where the contamination originated, then embedding this larger zoom region within the new root grid would solve this issue.

Similarly, one possible modification to the method presented here would be to use the trimmed white noise cube to generate a new zoom region in addition to the root grid, rather than copying the zoom region from the original IC. This would eliminate the discontinuity between the zoom region and the root grid cells, but would also likely lead to more significant changes in the galaxy properties. The hybrid IC method results in minor changes to the halo growth history and satellite distributions without changing the zoom region of the IC at all. Regenerating the zoom region would likely lead to much larger changes in galaxy properties. A major reason to use a hybrid IC is to facilitate code comparison with other groups using the same IC, and large changes in galaxy properties would eliminate this benefit.

\section{Code Speedup}

To test the code speedup provided by this method, we ran full hydrodynamic simulations of each realization of our IC using the latest version of the ART code. The code utilizes adaptive mesh refinement to reach high spatial resolution. It includes radiative transfer of ionizing and UV radiation from both stars and the extragalactic background. Radiative transfer is calculated using an improved version of the Optically Thin Variable Eddington Tensor method that minimizes numerical diffusion \citep{ngnedin14}. The ART code includes a non-equilibrium chemistry network that calculates the abundances of all species of hydrogen (H\,{\sc i}, H\,{\sc ii}, H$_2$) and helium (He\,{\sc i}, He\,{\sc ii}, He\,{\sc iii}), calibrated using observations in nearby galaxies \citep{ngnedin_kravtsov11}. A subgrid-scale model for numerically unresolved turbulence \citep{semenov_etal16} follows turbulent motions in the interstellar medium generated by stellar feedback. The most novel aspect of these simulations is the time-resolved modeling of star cluster formation \citep{li_etal17,li_etal18}. Growth of star clusters is terminated by their own feedback, allowing a self-consistent calculation of cluster masses. The resulting mass function of modeled clusters matches observations of young star clusters in nearby galaxies.

All of these simulations were run on the Stampede2 cluster at the Texas Advanced Computing Center using 8 Skylake nodes. The code version and setup of all runs were identical. The dark matter particle mass in the refined region is $1.57 \times 10^{5}\Msun$, with 47.5 million particles in this region. Within the refined region, we use adaptive mesh refinement to reach spatial resolution of $\approx 5$ pc (physical, not comoving), which is high enough to resolve giant molecular clouds. We refine cells if either their gas mass or dark matter mass exceeds a given threshold, which acts to keep all cells with roughly the same mass. Additionally, we use a Jeans length criterion, requiring at least 3 cells to resolve the local Jeans length. At $z\approx 10$, most of the cells in the refined region are at the original level or with one additional level of refinement, but densest regions within galaxies reach the maximum resolution of $\approx 5$ pc.

\begin{table}
    \centering
    \renewcommand{\arraystretch}{1.2}  
    \rowcolors{2}{white}{gray!15} 
    \begin{tabularx}{0.483\textwidth}{lXl>{\RaggedRight}X}
        \toprule
        Run & Time [hours] to $z\approx10$ & Speedup & Number of Cells at $z\approx10$ \\
        \midrule
        L1-128 & 80.0 & --- & 217,326,943 \\
        L2-256 & 57.5 & 1.39 & 219,741,152 \\
        L4-128 & 43.3 & 1.85 & 191,168,251 \\
        L4-256 & 43.6 & 1.83 & 204,306,243 \\
        \bottomrule
    \end{tabularx}
    \caption{The runtime of full hydro simulations to $z\approx10$ for different realizations of our IC using the ART code. These simulations were all done on Stampede2 with 8 Skylake nodes with identical setups (other than the IC). Here speedup is defined as the walltime of the original divided by the walltime of the other run.}
    \label{tab:runtime}
\end{table}

Table~\ref{tab:runtime} shows the walltime it took each of these simulations to reach $z\approx10$. All of the modified versions show substantial improvement. The L2-256 run is nearly 1.4 times faster than the original, while both L4 runs are more than 1.8 times faster. To determine the relative influence of different features of these realizations, we examine the particle number, cell number at $z \approx 10$, and the root grid cell size. While the cell number initially equals the particle number, runtime adaptive mesh refinement can be affected by slightly different growth of structure and stellar feedback.

First, the L2-256 and L4-128 runs have the same root grid cell size, but different particle and cell numbers. The L2-256 run took 1.33 times longer than the L4-128 run, consistent with the higher number of particles (1.28 times) and cells (1.15 times) in L2-256 run. This indicates that the walltime scales roughly with the particle and cell numbers, as expected. Second, we examine the effect of the root grid cell size. The L2-256 run has more cells and particles than the L1-128 run, but also 4 times smaller root grid cell size. It ran 1.4 times faster than the L1-128 run, indicating that smaller cell size did improve computational performance. However, smaller root grid cell sizes do not automatically lead to performance gains. In the L4-256 run the root grid cell size is half that of the L4-128 run, but the finer root grid results in more particles and cells, causing both runs to take nearly the same amount of time. Together, these results indicate that for the ART code, smaller root grid cell size does improve performance, but the increased particle and cell numbers required by finer root grids can negate some of these gains. We expect that load balancing gains enabled by the finer grid would become more important at later cosmic epochs.

\section{Summary}

We have presented a method for customizing the root grid surrounding the region of interest in a zoom-in cosmological simulation. We use the original white noise cube to produce structures that remain similar to the ones in the original IC, while allowing for customized box size and root grid. The modified root grid is combined with the original zoom region to produce an IC that minimally disturbs the galaxies in the zoom region. This method results in galaxies with similar properties to those in an unmodified simulation. The customization of the root grid can be tuned to maximize computational performance for a given code.

By reducing the cost of cosmological zoom-in simulations, this method will allow for more groups to run simulations using a common well-tested IC. This will enable more detailed code comparisons, as the confounding factor of groups using different ICs will be removed, and will allow the community to determine which aspects of galaxy formation are modeled most robustly.

\section*{Acknowledgements}
We thank Nick Gnedin and Oliver Hahn for fruitful discussions that informed our general approach and enabled us to adapt the \texttt{MUSIC} code for our purposes. This work was supported in part by the US National Science Foundation through grant 1909063.

\bibliographystyle{mnras}
\bibliography{ic,gc}

\end{document}